\documentclass[10pt,twocolumn,superscriptaddress,showpacs,amsmath,amssymb,osajnl,floatfix]{revtex4-1}
\usepackage{epsfig}
\usepackage{amsmath}
\usepackage{amssymb}
\usepackage{bm}
\usepackage{graphicx}
\usepackage{color}
\usepackage{graphicx,rotating}

\begin{document}
\title{Feasibility of electron cyclotron autoresonance acceleration by a short 
terahertz pulse}

\author{Yousef I. Salamin}
\affiliation{Max-Planck-Institut f\"{u}r Kernphysik, Saupfercheckweg 1,
69029 Heidelberg, Germany}
\affiliation{Department of Physics, American University of Sharjah, POB 26666,
Sharjah, United Arab Emirates (UAE)}

\author{Jian-Xing Li}
\affiliation{Max-Planck-Institut f\"{u}r Kernphysik, Saupfercheckweg 1,
69029 Heidelberg, Germany}

\author{Benjamin J. Galow\thanks{Present address: Gaisbergstra{\ss}e 61, 69115 Heidelberg (Germany).}}
\affiliation{Max-Planck-Institut f\"{u}r Kernphysik, Saupfercheckweg 1,
69029 Heidelberg, Germany}

\author{Christoph H. Keitel}
\affiliation{Max-Planck-Institut f\"{u}r Kernphysik, Saupfercheckweg 1,
69029 Heidelberg, Germany}

\date{\today}

\begin{abstract}
A vacuum autoresonance accelerator scheme for electrons, which employs terahertz
 radiation and currently available magnetic fields, is suggested. Based on
numerical simulations, parameter values, which could make the scheme
experimentally feasible, are identified and discussed.\\

\end{abstract}

\maketitle

\begin{figure}[t]
\includegraphics[width=8cm]{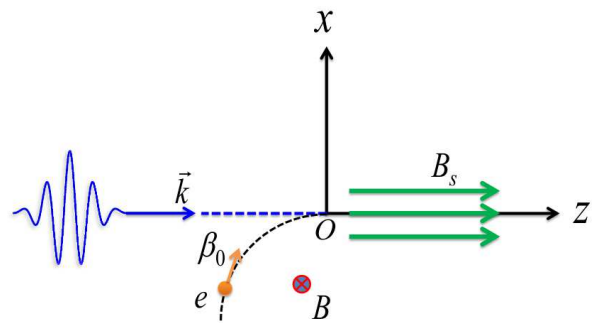}
\caption{(Color online) Schematic of a possible ALA setup. A magnetic field $B$
is used to bend the electron beam for collision with the THz pulse, which propagates along $+z$ (propagation vector $\vec{k}$). Front of the
pulse catches up with the electron precisely at the origin of coordinates O.
The magnetic field $\bm{B}_s=B_s\hat{k}$ is responsible for the electron
cyclotron motion.}
\label{fig1}
\end{figure}

The aim of this Letter is to investigate conditions for an electron vacuum 
autoresonance accelerator scheme that would employ circularly 
polarized terahertz (THz) radiation (or T-rays) and currently available
laboratory magnetic fields, with the hope of stimulating
future experiments. The system, subject of this Letter, is an electron (or 
electron bunch) injected in the common directions of radiation pulse propagation
 and an added uniform magnetic field (see Fig. \ref{fig1}).  

Autoresonance Laser Acceleration (ALA) of electrons has
a relatively long history \cite{milantev,bucksbaum1964,loeb1986,salamin2000,
singh1,singh2}.
Theoretical vacuum ALA studies
\cite{salamin2000,hirshfield2000} have shown that the
electron would gain a tremendous amount of energy from the laser field when the 
initial injection energy, the laser frequency $\omega$ and the external magnetic
field strength $B_s$, all
conspire to achieve resonance, or near-resonance. For axial injection, and
plane-wave fields, the resonance condition takes the form  
\begin{equation}\label{res}
 r \equiv \frac{\omega_c}{\omega_D}  = \frac{eB_{s}}{m\omega}
\sqrt{\frac{1+\beta_0}{1-\beta_0}} \to 1,
\end{equation}
where, $\omega_c = eB_{s}/m$ is the cyclotron frequency of the electron
around the lines of the static magnetic field, and $\omega_D$ is the
Doppler-shifted frequency of the radiation field, as seen by the electron. 
Furthermore, $m$ and $e$ are
the electron's mass and charge, respectively, and SI units have been used.
Also, $\beta_0$ is the speed with which the electron is injected,
scaled by the speed of light $c$. Autoresonance essentially means the electron
cyclotron frequency becomes comparable to the Doppler-shifted
frequency, seen by the electron, of the (circularly polarized) laser. When this happens, the velocity 
vector $\bm{\beta}$ of
the electron and the electric field vector $\bm{E}$ of the laser field
maintain the same angle during interaction, which leads, according to the
equation
\begin{equation}\label{rate}
	\frac{d\varepsilon}{dt}=-ec\bm{\beta}\cdot\bm{E},
\end{equation}
to synchronous energy gain by the particle from the radiation field. In Eq.
(\ref{rate}) the electron's relativistic energy is $\varepsilon=\gamma mc^2$,
where $\gamma=(1-\beta^2)^{-1/2}$. On resonance, the vectors $\bm{\beta}$ and
$\bm{E}$ gyrate, about the common direction of the magnetic field and laser
propagation, at the same frequency. According to Eq. (\ref{rate}), the field-to-electron 
energy transfer
rate is a maximum when $\bm{\beta}$ and $\bm{E}$ are antiparallel.

To the best of our knowledge, a vacuum autoresonace laser accelerator has not
been realized and none is
currently available and in operation. Plane-wave-based studies have 
shown that, for the scheme to work, a strong magnetic field needs to be
maintained over a long distance, which makes such a device both expensive and
prohibitively too long. Hope has been revived by recent investigations, in which
petawatt optical-frequency laser pulses, modeled most realistically 
\cite{galow2013} by Gaussian fields, have been employed. It has been shown that
electrons can gain over 10 GeV of energy in a magnetic field of strength
exceeding 50 T, and maintained constant over a distance
of several meters \cite{galow2013}. Unfortunately, even these conditions are at
present extremely difficult to realize.

In this Letter, the laser fields will be replaced by THz fields, and the
parameter space will be scanned in search of a parameter set which
may be more likely to be experimentally realized in the near future. 
According to the condition (\ref{res})
vacuum autoresonance may be achieved in a uniform magnetic field of a few tesla,
provided the radiation frequency is lowered by roughly two orders of magnitude. 
Lowering the frequency by two orders of magnitude from the optical domain lands
one in the THz
region of the electromagnetic spectrum, roughly in the range 0.3 -- 10 THz. 

\begin{figure}[t]
\includegraphics[width=8cm]{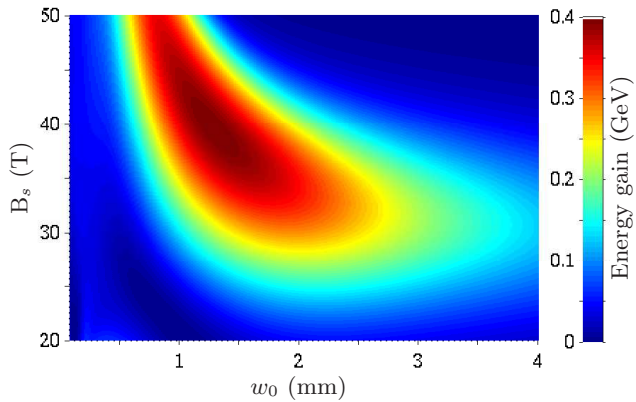}\\
\begin{picture}(0,0)(100,10)
\put(75,17){$w_0$ (mm)}
\put(211,60){\rotatebox{90}{Energy gain (GeV)}}
\put(-16,85){\rotatebox{90}{B$_s$ (T)}}
\end{picture}
\caption{(Color online) Contour plot of the energy gained by a single electron, 
injected with $\gamma_0=3$ for interaction with a single circularly polarized Gaussian pulse, as functions
of the magnetic field strength $B_s$ and the waist radius at focus $w_0$. The
pulse power is P = 100 TW, and its frequency is $f = 4$ THz ($\lambda = 75 ~\mu$m, period $T_0$ = 250 fs = FWHM).}
\label{fig2}
\end{figure}

\begin{figure}[t]
\vskip4mm
\includegraphics[width=8cm]{fig3a.eps}
\includegraphics[width=8cm]{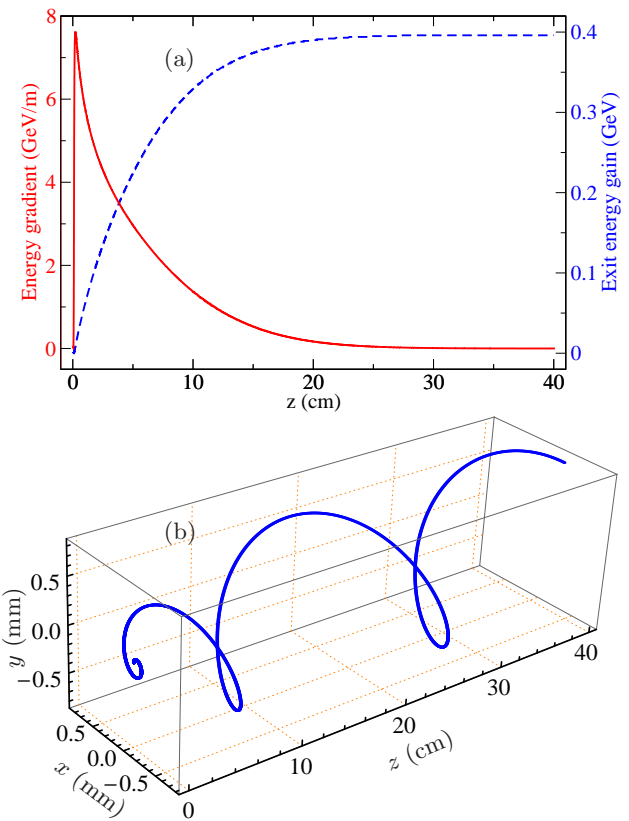}
\begin{picture}(0,0)(100,10)
\put(-75,297){(a)}
\put(-75,118){(b)}
\put(-118,31){\rotatebox{-37}{$x$ (mm)}}
\put(8,30){\rotatebox{23}{$z$ (cm)}}
\put(-136,69){\rotatebox{90}{$y$ (mm)}}
\end{picture}
\caption{(Color online) (a) Single electron energy gain (dashed blue) and energy gradient (solid red) as 
functions of its forward excursion distance. The parameters are those of
Fig. \ref{fig2}, in addition to $w_0=17\lambda\approx1.27$ mm, and $B_s\approx39.6$ T. (b) Actual trajectory of the 
electron under the conditions of (a).}
\label{fig3}
\end{figure}

Low-power THz radiation has been the subject of intense investigation for many 
years now, and applications in such fields as imaging, wireless communication and
remote sensing, have seen much progress recently \cite{armstrong,carr,neil,helm1,helm2,jepsen}. In
pondering
the idea of utilizing THz radiation in a vacuum ALA-like scheme, one must
immediately
come to grips with the need for high power. Currently considered the best source
for high-power THz radiation, a free-electron laser (FEL) can generate radiation
of only a few gigawatt (GW) power \cite{carr,neil}. Furthermore, promising candidates for 
generating more intense T-rays are the interactions of high intensity lasers 
with plasma or gas targets \cite{gopal,wang,chen}. Thus, to experimentally
realize a vacuum ALA-like THz scheme, the only remaining
(and admittedly non-trivial) challenge would be to develop sources of THz
radiation of power in the terawatt (TW) region and beyond, as will be
demonstrated  shortly. 

The single- and many-particle simulations in this Letter are all based on
solving  the
relativistic energy-momentum transfer equations 
\begin{equation}\label{motion}
 \frac{d\bm{\beta}}{dt} = \frac{e}{\gamma
mc}\left[\bm{\beta}(\bm{\beta}\cdot\bm{E})-\left(\bm{E}+c\bm{\beta}\times\bm{B}
\right)\right],
\end{equation} 
for the electron, subject to the adopted initial conditions of axial injection
with scaled energy $\gamma_0=(1-\beta_0^2)^{-1/2}$, together with the
assumption that the front of
the pulse catches up with the particle at $t=0$, precisely at the origin of
coordinates.  In
Eq. (\ref{motion}), $\bm{B}$ is the sum of the radiation magnetic field and
$\bm{B}_s$. Fields of the circularly polarized THz radiation are modeled
by those of a short Gaussian pulse \cite{li2014,salamin2007}.

In addition to the magnetic field strength $B_s$ and the THz frequency $\omega$,
a third
parameter, namely $\beta_0$ and, hence, the scaled injection energy of the
electron $\gamma_0$, plays a decisive role in determining 
the (exact, plane-wave-based) resonance
condition (\ref{res}). When a more
realistic model is adopted, such as that of a Gaussian beam or pulse,
detuning away from resonance results, and one is forced to search for
near-resonance by scanning the parameter space of Fig.
\ref{fig2}, for example \cite{galow2013}. 

For further discussion, we select from Fig. \ref{fig2} a parameter set 
which leads to near-autoresonance and, hence, high energy gain. With the
gain defined by 
\begin{equation}\label{G}
 G=(\gamma-\gamma_0)mc^2,
\end{equation} 
Fig. \ref{fig3}(a) shows that an electron exits with kinetic energy $K_{exit}\sim 396.2601$
MeV from interaction with a single-cycle, 100 TW pulse, over a distance of 
less than 40 cm, and in the presence of a
magnetic field of strength 39.6 T. Note that most of the energy is gained over
nearly the first 20 cm, and very little energy is gained beyond that point as the
electron interacts with the weak tail of the pulse and is ultimately left
behind. The energy gradient, or the 
gain per unit forward excursion distance,  
\begin{equation}
 \frac{dG}{dz} = -e\left(\frac{\bm{\beta}\cdot\bm{E}}{\beta_z}\right),
\end{equation} 
peaks over the first one centimeter and falls down to
zero very quickly. The peak gradient is almost 8 GeV/m, or about 80 times the
fundamental limit on the performance of conventional accelerators \cite{lee}. An {\it
average gradient}, found simply by dividing the total gain by the total
forward excursion during interaction with the pulse, is about 1 GeV/m, in this
example.

\begin{figure}[t]
\includegraphics[width=8cm]{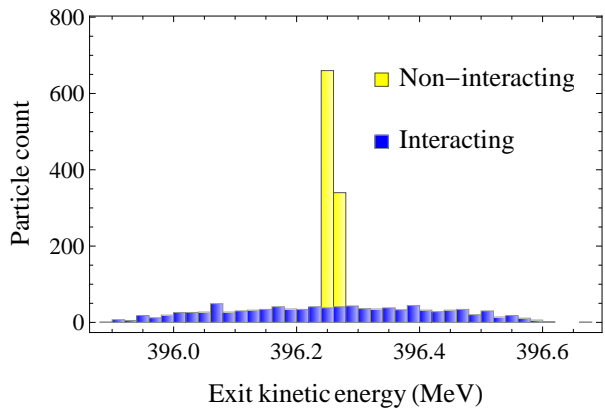}
\caption{(Color online) Exit kinetic energy distribution of a 1000  
ensemble of electrons. Interacting: Coulomb electron-electron interactions 
are turned on, and Non-interacting: Coulomb 
interactions are turned off. 
Electrons are assumed, initially, to be distributed randomly 
within a cylinder of radius 0.232 $\mu$m and height 4.642 $\mu$m, 
and centered at the origin of coordinates. Electron kinetic 
energy is, initially, distributed normally, with 
mean $\bar{K_0} = 1.022$ MeV and spread $\Delta K_0 = 0.1\%$.}
\label{fig4}
\end{figure}

Figure \ref{fig3}(b) shows the trajectory of the electron whose gain
and gradient are shown in (a). The trajectory is the expected
helix of increasing cross section \cite{salamin2000}. Note 
that the transverse dimensions of the helical trajectory are much
smaller than the longitudinal electron excursion, which makes the trajectory 
essentially linear.

The discussion so far has been limited to THz -- ALA of a single electron. Next,
we consider acceleration of an ensemble of 1000 electrons randomly distributed, 
initially, within a cylinder centered at the origin of coordinates, and whose
radius and height are 0.232 $\mu$m and 4.642 $\mu$m,
respectively. Initial kinetic
energy of the ensemble has a normal distribution of mean $\bar{K}_0 = 1.022$ MeV
and spread $\Delta K_0 = 0.1\%$. For the parameter set used in Fig.
\ref{fig3}, our simulations yield the exit energy
distributions displayed in the histograms of Fig. \ref{fig4}. From the data, one
gets $\bar{K}_{exit} = 396.2583\pm 0.0028$ MeV (or a spread of 0.0007 \%), with the
electron-electron
Coulomb interactions turned off, and $\bar{K}_{exit} = 396.2562\pm 0.1682$ MeV
(0.042 \%), when those
Coulomb interactions are properly taken into account. These
many-particle results agree quite well with our single-particle calculations
(Fig. \ref{fig3}, $K_{exit}\sim 396.2601$
MeV). Note also that the particle-particle interactions 
do result in a noticeable increase in the spread in exit energies. 
In absolute terms, however, the effect of incorporating the Coulomb interactions 
is small, due to the fact that the particle 
density of the initial ensemble is quite low ($\sim1.273\times10^{21}$
m$^{-3}$).

\begin{figure}[t]
\includegraphics[width=8cm]{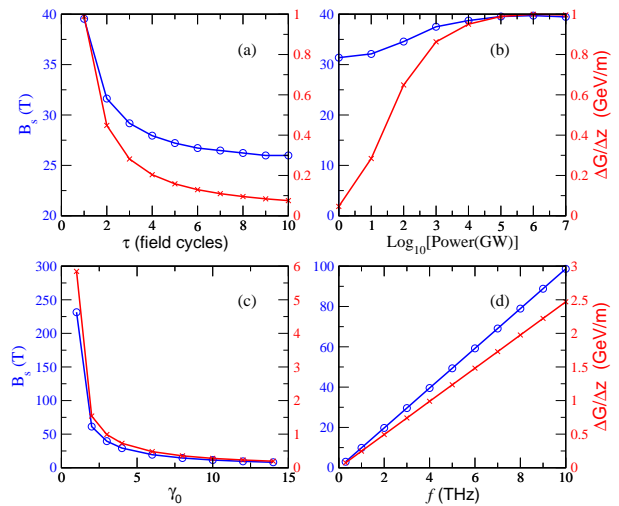}
\caption{(Color online) Near-resonance magnetic field dependence (blue and circles) upon: (a) the
THz pulse duration $\tau$ in units of the radiation field cycle $T_0$, (b) the THz
power,
(c) the scaled injection energy $\gamma_0$, and (d) the THz frequency. For each
data point in (a) -- (d) the average energy gradient in GeV/m is given (red and crosses).
Each data point is a result of calculations, along the lines 
of the work which led to Figs. \ref{fig2} and \ref{fig3}, 
and employing their (other) parameters. For example, in (a) 
$\gamma_0=3$, the power is 100 TW, $w_0=17~\lambda$, and $f=4$ THz, and so on.}
\label{fig5}
\end{figure}

Our focus in this Letter has been to demonstrate that THz -- ALA of electrons 
may be experimentally feasible with present-day technology, as far as
the needed magnetic field strengths are concerned. However, the example
presented above in some detail required the use of $B_s\sim 40$ T. 
This kind of field strength is available only at large and expensive facilities
\cite{singleton,lanl}. For weaker fields to be utilized, other parameters have
to be compromised. Values have to be picked for four more parameters which may
ultimately lead to energy gains and energy gradients more modest than has been
reported above. Figures \ref{fig5}(a) -- (d) show how the values of $B_s$,
corresponding to near-resonance, vary with the pulse duration, the power of the
THz radiation source, the scaled injection energy, and the THz frequency,
respectively. Shown in red also are the corresponding average energy gradients, 
defined in each case as the total exit energy gain divided by the total 
forward excursion distance, during interaction with the pulse. The results
displayed in Fig. \ref{fig5} emerge from calculations following 
along the same lines leading to Figs. \ref{fig2} and
\ref{fig3}, for each set of parameters separately. In most cases 
considered, the average energy gradients are much higher than 
the natural limit on performance of the conventional accelerators, 
namely, about 100 MeV/m. Note that in calculating these average 
gradients, the full excursion distance occurring during the full 
particle-field interaction time, is used. Considering 
that most of the energy is gained during interaction with only 
a small fraction of the excursion, leads to the conclusion 
that the energy gradients are, effectively, much higher
than is suggested by the averages reported in Fig. \ref{fig5}.

Figure \ref{fig5}(a) shows clearly that the use of currently
available laboratory magnetic fields in a THz -- ALA setup is feasible, but
the energy gradients that would be achieved fall to just a few times the
performance limit of conventional accelerators \cite{lee}. Recall 
that the example considered above (Fig. \ref{fig3}) made use of a 
one-cycle pulse. With several-cycle pulses, 
the resonance magnetic field strengths required 
fall already below 30 T, as shown in Fig. \ref{fig5}(a). The same
conclusions may be drawn from Fig. \ref{fig5}(c). 
However, the opposite trend exhibited in Figs.
\ref{fig5}(b) and (d) confirms earlier conclusions. 
Increasing the source power and frequency calls for 
the need for stronger resonance magnetic fields, 
and leads to high gains and
ultra-high energy gradients \cite{salamin2000,galow2013}. 
On-resonance dependence of $B_s$ upon the frequency $f$ 
is linear in Fig. \ref{fig5}(d), essentially like the 
plane-wave case (\ref{res}).

In conclusion, acceleration by cyclotron autoresonance 
of electrons in the simultaneous presence of THz fields 
and a uniform magnetic field of strength currently 
available for laboratory experiments, is feasible, 
provided a THz radiation source of TW power is also 
available. This is the case when the electrons are 
injected along the common directions of magnetic 
field and radiation propagation. According to 
Fig. \ref{fig5}(c), even acceleration from rest 
is possible, but the magnetic field strength 
needed for such a scheme is about 300 T, in this case.

\begin{acknowledgments}

YIS received partial support for this work from an American University of 
Sharjah Faculty Research Grant (FRG-III).

\end{acknowledgments}


\begin{thebibliography}{breitestes Label}
	\bibitem{milantev} V. P. Milant'ev, Phys.-Usp. {\bf 56}, 823 (2013).

\bibitem{bucksbaum1964} C. R. Roberts and S. J. Buchsbaum, Phys. Rev. {\bf 135}
  A381 (1964).

\bibitem{loeb1986} A. Loeb and L. Friedland, Phys. Rev. A {\bf 33}, 1828
(1986). 

\bibitem{salamin2000} Y. I. Salamin, F. H. M. Faisal, and C. H. Keitel, Phys.
Rev.  A {\bf 62}, 053809 (2000).

\bibitem{singh1} K. P. Singh, Phys. Rev. E {\bf 69}, 056410 (2004).

\bibitem{singh2} K. P. Singh, J. Opt. Soc. Am. B {\bf 23}, 1650 (2006).

\bibitem{hirshfield2000} J. L. Hirshfield, and C. Wang, Phys. Rev. E {\bf 61},
  7252 (2000).

\bibitem{galow2013} B. J. Galow, J-X. Li, Y. I. Salamin, Z. Harman, and C. H.
  Keitel, Phys. Rev. ST-AB {\bf 16}, 081302 (2013).

\bibitem{armstrong} C. M. Armstrong, IEEE Spectrum, 
  \url{http://spectrum.ieee.org/aerospace/military/the-truth-about-terahertz}.

\bibitem{carr} G. L. Carr, M. C. Martin, W. R. McKinney, K. Jordan, G. R. Neil,
  and G. P. Williams, Nature {\bf 420}, 153 (2002).

\bibitem{helm1} M. Walther, B. Fischer, M. Schall, H. Helm, and P. U. Jepsen, Chem. Phys. Lett.  {\bf 332}, 389 (2000).

\bibitem{helm2} A. Bitzer, H. Merbold, A. Thoman, T. Feurer, H. Helm, and M. Walther, Opt. Express {\bf 17}, 3826 (2009).

\bibitem{jepsen} P. U. Jepsen, D. G. Cooke, and M. Koch, Las. \& Phot. Rev. {\bf 5}, 124 (2011).

\bibitem{neil} G. R. Neil, J. Infrared Milli. Terahz. Waves {\bf 35}, 5 (2014).

\bibitem{gopal} A. Gopal, S. Herzer, A. Schmidt, P. Singh, A. Reinhard, W. Ziegler, D. Br\"ommel, A. Karmakar, P. Gibbon, U. Dillner, T. May, H-G Meyer, and G. G. Paulus, Phys. Rev. Lett. {\bf 111}, 074802(2013).

\bibitem{wang} W.-M. Wang, Z.-M. Sheng, H.-C. Wu, M. Chen, C. Li, J. Zhang, and K. Mima, Opt. Express {\bf16}, 16999-17006 (2008).

\bibitem{chen} Z.-Y. Chen, X.-Y. Li, and Y. Wei, Phys. Plasmas {\bf 20}, 103115 (2013).

\bibitem{li2014} J.-X. Li, K. Z. Hatsagortsyan, and C. H. Keitel, Phys. Rev.
	Lett. {\bf 113}, 044801 (2014).

\bibitem{salamin2007} Y. I. Salamin, Appl. Phys. B {\bf 86}, 319 (2007).

\bibitem{lee} S. Y. Lee, Accelerator Physics, 2nd Ed. (World Scientific, Singapore, 2004).

\bibitem{singleton} J. Singleton, C. H. Mielke, A. Migliori, G. S. Boebinger,
	and A. H. Lacerda, Physica B{\bf 346}, 614 (2004).

\bibitem{lanl} See \url{http://www.lanl.gov/orgs/mpa/nhmfl/}, and
	\url{http://www.lanl.gov/orgs/mpa/nhmfl/60TLP.shtml}.
\end{thebibliography}
\end{document}